\documentclass[aps,pre,preprint,floatfix,twoside,tightenlines,showpacs,
showkeys]{revtex4}
\usepackage{amsmath,graphicx}
\begin{document}
%
\title{Field theoretical representation of classical statistical mechanics. I. Wave-vector space}
\author{A.Yu.~Zakharov}\email[E-mail: ]{Anatoly.Zakharov@novsu.ru}
\author{E.V.~Sokolova} \email[E-mail: ]{lenux999@yandex.ru}
\affiliation{Novgorod State University, Novgorod the Great, 173003, Russia}
%
%
\begin{abstract}
Thermodynamic equivalence between classical many-body system and some auxiliary nonlinear auxiliary field is proved. Connection between Hamiltonians of the many-body system and the auxiliary field is derived.
\end{abstract}
%
\maketitle

\section{Introduction}

There are a few different starting points for setting of the statistical mechanics problems. These starting points are canonical, microcanonical, grand canonical and other ensembles. In general, these ensembles are not equivalent. Moreover, nonequivalence of the ensembles occurs for most interesting problems, such as new phases nucleations, statistics of small systems, metastable states etc. Therefore, statistical mechanics in its present state is not a well-posed theory. 

From this standpoint, it is interesting to develop the general methods that can be used to different ensembles so as to different models of statistical mechanics. The problem of the canonical configuration integral calculation exists in every ensemble of statistical mechanics with some variations: for canonical and grand canonical ensembles the configuration integral depends on the {\em real} temperature, whereas for microcanonical ensemble this  integral depends on the {\em imaginary} temperature~\cite{Zak0}. The main difficulty in the partition function calculation caused the atomic coordinates entanglements in the integrand. Factorization of the integrand on the atomic coordinates by means of the functional integral representation solves the entanglements problem. But instead of that, the new problem of the non-Gaussian functional integral calculation occurs. 

Probably, for the first time the notion of the functional integration was introduced by Wiener~\cite{Wie}. In connection with statistical mechanics the functional integration method was initialized by Zubarev~\cite{Zub} and Edwards~\cite{Edw}. Some mathematical aspects  of the functional integration method, including approximate calculation of functional integrals,  are systematized by Egorov, Zhidkov and Lobanov~\cite{Ego}.

Functional integrals include integration over some set of the auxiliary functions. In accordance with domain of definition of the auxiliary functions, there are two basic variants of the functional integral representations for classical partition functions in classical statistical mechanics.
\begin{enumerate}
    \item Functional integration over the functions defined in the real space~\cite{Edw,Iva1,Iva2};
    \item Functional integration over the functions defined in the wave-vectors space~\cite{Zub,Ami,Par,Zak1}.
\end{enumerate}
In both of these variants the partition functions or the configuration integrals calculations reduced to the non-Gaussian functional integral evaluations. 

Functional integration approach has essential advantages over the direct analysis of the configuration integrals. In particular, this method makes possible
\begin{enumerate}
    \item  avoidance of the combinatorial problems related to the Mayer expansion~\cite{Iva2,Par} and some other similar problems;
    \item analysis of some simple models statistical mechanics such as the systems with non-negative Fourier-transforms of the interatomic potentials \cite{Zak2,Efi,Bae1,Bae2}; 
\end{enumerate}
However, the auxiliary field in the functional integrals is an interesting object by itself. It gives a new way of the statistical mechanical problems formulation. 

The object of this paper is the formulation of the new variant of the statistical mechanical problems setting. This variant based on the functional integration method over the auxiliary functions defined in the wave-vectors space.

\section{Functional integral representation in wave-vector space}

The generating functional of a system of $N$ particles, interacting via two-body potential~$v\left(\textbf{r}\right)$ in presence of an external field $\varphi\left(\textbf{r}\right)$, after integration with respect to momentum has the known form
\begin{equation}\label{Stat}
\begin{array}{r}
{\displaystyle
Z{\left\{  \varphi(\mathbf{r})\right\} } =\frac{V^{N}}{N!\lambda^{DN}} \idotsint\limits_{(V)}\ \left(\prod^{N}_{s=1}\frac{d^{D}R_{s}}{V}\right)\exp\left( -\beta\sum\limits_{s=1}^N \varphi(\textbf{R}_s)\right)}\\ \\
{\displaystyle
\times
\exp\left(-\frac{\beta}{2}\sum\limits_{\stackrel{\scriptstyle s,s'=1,}{s\neq s'}  }^{N} v\left(\textbf{R}_{s} - \textbf{R}_{s'}\right)\right)}{,}
\end{array}
\end{equation}
where $\lambda=\left(2\pi\hbar/m k_{B}T\right)^{1/2}$ is de Broglie thermal wavelength, $\beta=1/k_{B}T$ is reciprocal temperature, $V$ is the system volume, $D$ is the space dimensionality, $\hbar$ and $k_{B}$ are the Planck and Boltzmann constants, respectively.

The first exponent in the integrand separates into one-atom multipliers in contrast of the second exponent. Suppose that the central interatomic potential $v\left(\textbf{r}\right)$ assumes expansion into the Fourier series. Then the pairwise interactions energy of the particles in the system can be written as  
\begin{equation}
\label{pot1}
\begin{array}{r}
{\displaystyle 
\frac{1}{2} \sum\limits_{\stackrel{\scriptstyle s,s'=1,}{s\neq s'} }^N v(\textbf{R}_s - \textbf{R}_{s'}) = \frac{N}{2}\left(n\tilde{v} \left(0\right) - v\left(0\right)\right) }\\ \\
{\displaystyle  + \frac{1}{2V}\sum\limits_{\textbf{k}\in \Omega^+}v^{+}\left(\textbf{k}\right)\left[C^{2}\left(\textbf{k}\right) + S^{2}\left(\textbf{k}\right)\right]  
 - \frac{1}{2V}\sum\limits_{\textbf{k}{'}\in\Omega^{-}} v^{-}\left(\textbf{k}{'}\right)\left[C^{2}\left(\textbf{k}{'}\right) + S^{2}\left(\textbf{k}{'}\right)\right]},
\end{array}
\end{equation}
where $C\left(\mathbf{k}\right)$ and $S\left(\mathbf{k}\right)$ are the collective variables
\begin{equation}
\label{cos}
C\left(\mathbf{k}\right)= \sum\limits^{N}_{s=1}\cos\left(\mathbf{k}\textbf{R}_s\right){;} \quad S\left(\mathbf{k}\right)= \sum\limits^{N}_{s=1}\sin\left(\mathbf{k}\textbf{R}_s\right){,}
\end{equation}
summation over $\mathbf{k}$ and $\mathbf{k}'$ in (\ref{pot1}) does not contain the point $\mathbf{k}=0$, the term ${-}\frac{N}{2}v\left(0\right)$ compensates the summands with $s = s{'}$ in the right hand side of (\ref{pot1}), $\Omega^{\pm}$ denote the subsets of the wave vector space in which the Fourier transform $\tilde{v}\left(\textbf{k}\right)$ of the interatomic potential is positive and negative, respectively, 
\begin{equation}\label{v-pm}
\left\{
\begin{array}{l}
    {\displaystyle v^+\left(\mathbf{k}\right)=+\tilde{v},\quad \mathbf{k}\in\Omega^+ ;}\\
{\displaystyle v^-\left(\mathbf{k}\right)=-\tilde{v},\quad  \mathbf{k}\in\Omega^-, }
\end{array}
\right.
\end{equation}
$v^{\pm}(\mathbf{k})>0$, $n=N/V$.

The second exponent in the integrand of~(\ref{Stat}) as function of the collective variables has the following form:
\begin{equation}\label{Z-C-S}
\begin{array}{r}
    {\displaystyle \exp\left(-\frac{\beta}{2}\sum\limits_{\stackrel{\scriptstyle s,s'=1,}{s\neq s'}  }^{N} v\left(\textbf{R}_{s} - \textbf{R}_{s'}\right)\right) = \exp {\left( \frac{\beta N}{2} \left[v(0) - n\tilde{v}(0) \right] \right)}}\\ \\
\times {\displaystyle  \left[ \prod_{\mathbf{k}\in \Omega^+} e^{ \left( -\frac{\beta v^+(\mathbf{k})}{2V} \left\{ C^2(\mathbf{k}) + S^2(\mathbf{k}) \right\} \right)} \right]\, \left[ \prod_{\mathbf{k}'\in \Omega^-} e^{ \left( \frac{\beta v^-(\mathbf{k}')}{2V} \left\{ C^2(\mathbf{k}') + S^2(\mathbf{k}') \right\} \right)} \right] }.
\end{array}
\end{equation}

Applying the Stratonovich-Hubbard transformation
\begin{equation}\label{SH}
\left\{
\begin{array}{c}
    {\displaystyle e^{-\frac{B^2}{2A}}=\sqrt\frac{A}{2\pi} \int\limits_{-\infty}^{+\infty}e^{-\frac{A}{2}x^{2}+iBx}\,dx,}\\
{\displaystyle e^{+\frac{B^2}{2A}}=\sqrt\frac{A}{2\pi} \int\limits_{-\infty}^{+\infty}e^{-\frac{A}{2}x^{2}+Bx}\,dx}
\end{array}
\right.
\end{equation}
for each of the exponents in~(\ref{Z-C-S})
\begin{equation}\label{e1}
\begin{array}{r}
{\displaystyle e^{ \left( -\frac{\beta v^+(\mathbf{k})}{2V} \left\{ C^2(\mathbf{k}) + S^2(\mathbf{k}) \right\} \right)} = \iint\limits_{-\infty}^{+\infty} \frac{\beta m \omega^{2} (\mathbf{k})\, dx^{+}(\mathbf{k})\, dy^{+} (\mathbf{k})}{2\pi}\ e^{ -\frac{\beta m \omega^{2}(\mathbf{k})\left(\left[x^{+} \left(\mathbf{k}\right)\right]^{2} + \left[y^{+}\left(\mathbf{k}\right)\right]^{2}\right)}{2}} }\\
\times\ {\displaystyle e^{i\beta \sqrt{\frac{m v^+(\mathbf{k})}{V}} \omega(\mathbf{k})  \left[ x^+(\mathbf{k}) C(\mathbf{k}) + y^+(\mathbf{k}) S(\mathbf{k}) \right] }\, ,  }
\end{array}
\end{equation}
\begin{equation}\label{e2}
\begin{array}{r}
{\displaystyle e^{ \left( \frac{\beta v^-(\mathbf{k}')}{2V} \left\{ C^2(\mathbf{k}') + S^2(\mathbf{k}') \right\} \right)} = \iint\limits_{-\infty}^{+\infty} \frac{\beta m \omega^{2} (\mathbf{k}')\, dx^{-}(\mathbf{k}')\, dy^{-} (\mathbf{k}')}{2\pi}\ e^{ -\frac{\beta m \omega^{2}(\mathbf{k}')\left(\left[x^{-}(\mathbf{k}')\right]^{2} + \left[y^{-}(\mathbf{k}')\right]^{2}\right)}{2}} }\\
\times\ {\displaystyle e^{\beta \sqrt{\frac{m v^-(\mathbf{k}')}{V}} \omega(\mathbf{k}')  \left[ x^-(\mathbf{k}') C(\mathbf{k}') + y^-(\mathbf{k}') S(\mathbf{k}') \right] }\, ,  }
\end{array}
\end{equation}
we obtain the following functional integral representation for $Z{\left\{  \varphi(\mathbf{r})\right\}}$:
\begin{equation}
\label{Z}
\begin{array}{r}
{\displaystyle 
Z{\left\{  \varphi(\mathbf{r})\right\} } = \frac{V^{N}}{N!\lambda^{DN}}e^{\left( \frac{\beta N}{2} \left[v(0) - n\tilde{v}(0) \right] \right)}
}\\
\times\ {\displaystyle 
\idotsint\limits_{-\infty}^{+\infty} \left(\prod_{\mathbf{k}\in\Omega^{+}} \frac{\beta\, m\, \omega^{2}\left(\mathbf{k}\right)\,dx^{+}\left(\mathbf{k}\right)dy^{+} \left(\mathbf{k}\right)}{2\pi}\right) \left(\prod_{\mathbf{k}'\in\Omega^{-}} \frac{\beta\, m\, \omega^{2}\left(\mathbf{k}'\right)dx^{-}\left(\mathbf{k}'\right)dy^{-} \left(\mathbf{k}'\right)}{2\pi}\right)
}\\
{\displaystyle 
\times {\exp\left(- \beta \left\{ \sum_{\mathbf{k}\in\Omega^+}\frac{ m\, \omega^{2}\left(\mathbf{k}\right)\left(\left[x^{+} \left(\mathbf{k}\right)\right]^{2} + \left[y^{+}\left(\mathbf{k}\right)\right]^{2}\right)}{2} \right\}\right)}
}\\
{\displaystyle 
\times {\exp\left(- \beta \left\{ \sum_{\mathbf{k}'\in\Omega^-}\frac{ m\, \omega^{2}\left(\mathbf{k}'\right)\left(\left[x^{-} \left(\mathbf{k}'\right)\right]^{2} + \left[y^{-}\left(\mathbf{k}'\right)\right]^{2}\right)}{2} \right\}\right)}
}\\ {\displaystyle \times
\left[\mathcal{F}_{1}\left(x^{\pm}\left(\mathbf{k}\right),\ y^{\pm}\left(\mathbf{k}\right),\ \left\{\varphi\left(\mathbf{r}\right)\right\}\right)\right]^{N}}\, ,
\end{array}
\end{equation}
where 
\begin{equation}
\label{Fu}
\begin{array}{r}
{\displaystyle 
\mathcal{F}_{1}\left(x^{\pm}\left(\mathbf{k}\right),\ y^{\pm}\left(\mathbf{k}\right),\ \left\{\varphi\left(\mathbf{r}\right)\right\}\right) = \int\limits_{(V)}\frac{d\mathbf{r}}{V}\, e^{{-}\beta\varphi\left(\mathbf{r}\right)}} \\
{\displaystyle 
\times\exp\left\{i\beta\sum_{\mathbf{k}\in\Omega^{+}} \sqrt{\frac{m v^{+}\left( \mathbf{k}\right)}{V}}\ \omega \left(\mathbf{k}\right) \left[x^{+}\left(\mathbf{k}\right) \cos\left(\mathbf{k}\mathbf{r}\right) + y^{+}\left(\mathbf{k}\right) \sin\left(\mathbf{k}\mathbf{r}\right)\right]\right\}} \\
{\displaystyle \times
\exp\left\{\beta\sum_{\mathbf{k}'\in\Omega^{-}}\sqrt{\frac{m v^{-}\left(\mathbf{k}'\right)}{V}}\ \omega\left(\mathbf{k}'\right) \left[x^{-}\left(\mathbf{k}'\right)\cos \left(\mathbf{k}'\mathbf{r}\right) + y^{-}\left(\mathbf{k}'\right) \sin\left(\mathbf{k}'\mathbf{r}\right)\right]\right\}},
\end{array}
\end{equation}
$\omega\left(\mathbf{k}\right)$ is an arbitrary positive function of the wave-vector with the dimensionality of the circular frequency, $m$ is an arbitrary positive parameter with the dimensionality of mass, and  $x^{\pm}\left(\mathbf{k}\right)$, $y^{\pm}\left(\mathbf{k}\right)$ are auxiliary  variables as a result of Stratonovich-Hubbard transformation.

Functional $ \mathcal{F}_{1}\left(x^{\pm}\left(\mathbf{k}\right),\ y^{\pm}\left(\mathbf{k}\right),\ \left\{\varphi\left(\mathbf{r}\right)\right\}\right) $ has a very simple physical interpretation: it is the configuration integral for one particle in {\em complex-valued} external field $\Psi\left(\textbf{r}\right)$:
\begin{equation}
\label{psi}
\begin{array}{r}
{\displaystyle 
\Psi\left(\mathbf{r}\right)=\varphi\left(\mathbf{r}\right) - i\sum_{\mathbf{k}\in\Omega^{+}}\sqrt{\frac{mv^{+} \left(\mathbf{k}\right)}{V}}\omega\left(\mathbf{k}\right) \left[x^{+}\left(\mathbf{k}\right) \cos\left(\mathbf{k}\mathbf{r}\right) + y^{+}\left(\mathbf{k}\right) \sin\left(\mathbf{k}\mathbf{r}\right)\right]}\\ 
{\displaystyle -
\sum_{\mathbf{k}{'}\in\Omega^{-}} \sqrt{\frac{mv^{-}\left(\mathbf{k}{'}\right)}{V}} \omega\left(\mathbf{k}{'}\right) \left[x^{-}\left(\mathbf{k}{'}\right)\cos \left(\mathbf{k}{'}\mathbf{r}\right) + y^{-}\left(\mathbf{k}'\right)\sin\left(\mathbf{k}'\mathbf{r}\right)\right]}.
\end{array}
\end{equation}
This complex-valued external field consists of two parts: the ``true'' external field $\varphi\left(\mathbf{r}\right)$ and some 
artificial complex field caused by exclusion of the interatomic interactions   and represented in the form of the Fourier series with 
its coefficients $\sqrt{\frac{mv^{\pm}\left(\mathbf{k}\right)}{V}}\, \omega\left(\mathbf{k}\right)x^{\pm}\left(\mathbf{k}\right)$, $\sqrt{\frac{mv^{\pm}\left(\mathbf{k}\right)}{V}}\, \omega\left(\mathbf{k}\right)y^{\pm}\left(\mathbf{k}\right)$. 
The expression $\mathcal{F}_{1}\left(x^{\pm}\left(\mathbf{k}\right), y^{\pm}\left(\mathbf{k}\right), \left\{\varphi\left(\textbf{r}\right)\right\}\right)^{N}$ is a configuration integral of a classical ideal gas in the external 
field $\Psi\left(\textbf{r}\right)$.

In accordance with (\ref{Z}), {\bf any classical system of interacting particles is statistically equivalent to an ideal gas in the complex-valued external field with Gaussian distribution of its the Fourier coefficients}.

There are a few variants for functional integral representations like~(\ref{Z}) of the canonical and grand canonical partition functions in statistical mechanics. These representations can be used for analysis of some concrete models~\cite{Zak2,Efi,Bae1,Bae2}. But, there are not any general methods of the functional integrals calculation excepting Gaussian integrals. Therefore, most of the results, related to the functional integral representations of the partitions functions in statistical mechanics, are obtained by the following ways:
\begin{enumerate}
    \item The saddle point method. This method is consistent if $\Omega^- = \emptyset$ and leads to Gaussian functional integral. The saddle point method relates to the mean field approximation. Otherwise (i.e. if $\Omega^- \not= \emptyset$), the location of the saddle point in space of the auxiliary field components depends on temperature and Gaussian approximation is inadequate. This problem has not solution yet. 
    \item The expansions of the functional integral in powers of the auxiliary fields $x^{\pm}(\mathbf{k})$, $y^{\pm}(\mathbf{k})$ with summation  of some finite number of the terms in this series. In this connection, there is no results on this series convergences.
\end{enumerate}

\section{Hamiltonian of Auxiliary Field}

The expression (\ref{Z}) can be interpreted as a statistical integral of the system of interacting oscillators. Substitute the identity 
\begin{equation}
\label{1}
\frac{\beta}{2\pi m}\int\int dp_{x}dp_{y}\exp\left[{-}\beta\left(\frac{p^{2}_{x}}{2m} + \frac{p^{2}_{y}}{2m}\right)\right] = 1,
\end{equation}
into~(\ref{Z}) and obtain
\begin{equation}
\label{ZZ}
\begin{array}{r}
{\displaystyle 
Z{\left\{  \varphi(\mathbf{r})\right\} } = \frac{V^{N}}{N!\lambda^{DN}} e^{\left( \frac{\beta N}{2} \left[v(0) - n\tilde{v}(0) \right] \right)}}\\ \\
{\displaystyle\times
\idotsint\limits_{-\infty}^{+\infty}\left(\prod\limits_{\mathbf{k}\in\Omega^{+}}\left[\frac{\beta\omega \left(\mathbf{k}\right)}{2\pi}\right]^{2}dx^{+}\left(\mathbf{k} \right)dp^{+}_{x}\left(\mathbf{k}\right)dy^{+}\left(\mathbf{k} \right)dp^{+}_{y}\left(\mathbf{k}\right)\right)}\\ \\
{\displaystyle\times
\left(\prod\limits_{\mathbf{k}'\in\Omega^{-}}\left[\frac{\beta\omega \left(\mathbf{k}'\right)}{2\pi}\right]^{2}dx^{-}\left(\mathbf{k}{'} \right)dp^{-}_{x}\left(\mathbf{k}'\right)dy^{-}\left(\mathbf{k}' \right)dp^{-}_{y}\left(\mathbf{k}'\right)\right)}\\ \\
{\displaystyle \times
\exp\left\{{-}\beta\sum\limits_{\mathbf{k}\in\Omega^{+}}\left(\frac{\left( \left[p^{+}_{x}\left(\mathbf{k}\right)\right]^{2} + \left[p^{+}_{y}\left(\mathbf{k}\right)\right]^{2}\right)}{2m} + \frac{m\omega^{2}\left(\mathbf{k}\right)\left(\left[x^{+}\left(\mathbf{k} \right) \right]^{2} + \left[y^{+}\left(\mathbf{k}\right)\right]^{2}\right)}{2}\right)\right\} }\\ \\
{\displaystyle\times
\exp\left\{{-}\beta\sum\limits_{\mathbf{k}'\in\Omega^{-}}\left(\frac{\left( \left[p^{-}_{x}\left(\mathbf{k}'\right)\right]^{2} + \left[p^{-}_{y}\left(\mathbf{k}'\right)\right]^{2}\right)}{2m} + \frac{m\omega^{2}\left(\mathbf{k}'\right)\left(\left[x^{-}\left( \mathbf{k}'\right)\right]^{2} + \left[y^{-}\left(\mathbf{k}'\right)\right]^{2}\right)}{2}\right)\right\}}\\ \\
{\displaystyle\times
\left[\mathcal{F}_{1}\left(x^{\pm}\left(\mathbf{k}\right),\ y^{\pm}\left(\mathbf{k}\right),\ \left\{\varphi\left(\mathbf{r}\right)\right\}\right)\right]^{N}}.
\end{array}
\end{equation}
This integral represents the classical partition function of a system described by the Hamiltonian $\tilde{\mathcal{H}}$:
\begin{equation}
\label{Ham}
\begin{array}{r}
{\displaystyle
\tilde{\mathcal{H}} = \sum\limits_{\mathbf{k}\in\Omega^{+}} \left[\frac{\left[p^{+}_{x}\left(\mathbf{k}\right)\right]^{2}}{2m} + \frac{m\omega^{2} \left(\mathbf{k}\right) \left(x^{+}\left(\mathbf{k}\right)\right)^{2}}{2} +
\frac{\left[p^{+}_{y}\left(\mathbf{k}\right)\right]^{2}}{2m} + \frac{m\omega^2\left(\mathbf{k}\right)\left(y^{+} \left(\mathbf{k}\right)\right)^{2}}{2}\right] }\\ \\
{\displaystyle+ 
\sum\limits_{\mathbf{k}'\in\Omega^{-}} \left[\frac{\left[p^{-}_{x}\left(\mathbf{k}'\right)\right]^{2}}{2m} + \frac{m\omega^{2}\left(\mathbf{k}'\right) \left(x^{-}\left(\mathbf{k}'\right)\right)^{2}}{2} + 
\frac{\left[p^{-}_{y}\left(\mathbf{k}'\right)\right]^{2}}{2m} + \frac{m\omega^2\left(\mathbf{k}'\right) \left(y^{-}\left(\mathbf{k}'\right)\right)^{2}}{2}\right]}\\ \\
{\displaystyle -
\frac{N}{\beta}\ln{\mathcal{F}_{1}\left(x^{\pm}\left(\mathbf{k}\right),\ y^{\pm}\left(\mathbf{k}\right),\ \left\{\varphi\left(\mathbf{r}\right)\right\}\right)}}.
\end{array}
\end{equation}
The multipliers $\frac{2\pi}{\beta\omega(\mathbf{k})}$ in the integrand of~(\ref{ZZ}) have the dimensionality of the Planck constant and play the role of an elementary cell volume in related phase space.

Hamiltonian (\ref{Ham}) depends on the generalized coordinates $x^{\pm}\left(\mathbf{k}\right)$,  $y^{\pm}\left(\mathbf{k}\right)$ and the generalized momenta $p^{\pm}_{x}\left(\mathbf{k}\right)$, $p^{\pm}_{y}\left(\mathbf{k}\right)$. In addition, $\tilde{\mathcal{H}}$ contains the temperature as parameter in the term $\mathcal{F}_{1}\left(x^{\pm} \left(\mathbf{k}\right),\ y^{\pm}\left(\mathbf{k}\right),\ \left\{\varphi\left(\mathbf{r}\right)\right\}\right)$. 

Thus, {\bf the partition functions for a classical many-body system and for the auxiliary field with Hamiltonian~(\ref{Ham}) are identical}.

\section{conclusion}
In accordance with relation~(\ref{ZZ}), the problem of obtaining the thermodynamic properties of many-body system is equivalent to the analogical problem of a system described by Hamiltonian~(\ref{Ham}). This Hamiltonian contains the summands corresponding to a system non-interacting classical harmonic oscillators (the first and the second lines in formula~(\ref{Ham}));  the third lines of~(\ref{Ham})  contains interactions between these fictitious oscillators including their anharmonicities. 

This Hamiltonian leads to some non-linear equation of the auxiliary field motion. The analysis of the auxiliary field evolution on base of these equations can be used for the statistical properties of the auxiliary fields by analogy with molecular dynamics method.


\begin{thebibliography}{99}


\bibitem{Zak0} {\it A.Yu.~Zakharov.} Techn. Phys. Lett., 2003. {\bf 29}. 791.

\bibitem {Wie} {\it N.~Wiener.} J. Math. and Phys. Sci., 1923. V. {\bf 2}. 131.

\bibitem{Zub} {\it D.N.~Zubarev.} Dokl. Akad. Nauk SSSR, 1954. {\bf 95}. No.4. 757.
%
\bibitem{Edw} {\it S.F.~Edwards.}  Phil. Mag., 1959. {\bf 46}. 1171.
%
\bibitem{Ego} {\it A.D.~Egorov, E.P.~Zhidkov, Yu.Yu.~Lobanov.} Introduction into Theory and Applications of Functional Integration. Moscow: Fizmatlit, 2006. (In Russian).
%
\bibitem{Iva1} {\it Yu.M.~Ivanchenko, A.A.~Lisyansky.} Phys. Lett. A, 1983. {\bf 98}. 115. 

\bibitem{Iva2} {\it Yu.M.~Ivanchenko, A.A.~Lisyansky.} Physics of Critical Fluctuations. N.Y.: Springer, 1995. 

\bibitem{Ami} {\it D.J.~Amit.} Field Theory, the Renormalization Group, and Critical Phenomena. N.Y.: McGraw-Hill, 1978.
%
\bibitem{Par} {\it G.~Parisi.} Statistical Field Theory. Reading, Massachusetts: Addison-Wesley, 1988. 
%
\bibitem{Zak1} {\it A.Yu.~Zakharov.}  Phys. Lett. A. 1990. {\bf 147}. 442. 
%
\bibitem{Zak2} {\it A.Yu.~Zakharov, I.K.~Loktionov.} Theor. Math. Phys. 1999. {\bf 119}. 532.
%
\bibitem{Efi} {\it G.V.~Efimov, E.A.~Nogovitsin.} Physica A. 1996. {\bf 234}. 506.
%
\bibitem{Bae1} {\it S.A.~Baeurle, R.~Marto\v{n}\'ak, M.~Parrinello.} J. Chem. Phys. 2002. {\bf 117}. 3027.
%
\bibitem{Bae2} {\it S.A.~Baeurle, G.V.~Efimov, E.A.~Nogovitsin.} J. Chem. Phys. 2006. {\bf 124} 224110.
%
%

\end{thebibliography}
\end{document}